\documentclass[a4paper]{spie}  

\usepackage{amsmath,amsfonts,amssymb}
\usepackage{graphicx}
\usepackage[colorlinks=true, allcolors=blue]{hyperref}

\usepackage{eurosym}

\title{HiPERCAM: A high-speed, quintuple-beam CCD camera for the study
  of rapid variability in the Universe}

\author[a,b]{Vikram S. Dhillon}
\author[c]{Thomas R. Marsh}
\author[d]{Naidu Bezawada}
\author[d]{Martin Black}
\author[a]{Simon Dixon}
\author[a]{Trevor Gamble}
\author[d]{David Henry}
\author[a]{Paul Kerry}
\author[a]{Stuart P. Littlefair}
\author[d]{David W. Lunney}
\author[e]{Timothy Morris}
\author[e]{James Osborn}
\author[e]{Richard W. Wilson}
\affil[a]{Department of Physics and Astronomy, University of
  Sheffield, Sheffield S3 7RH, UK}
\affil[b]{Instituto de Astrof\'{i}sica de Canarias, E-38205 La Laguna,
  Tenerife, Spain}
\affil[c]{Department of Physics, University of Warwick, Coventry CV4
  7AL, UK}
\affil[d]{UK Astronomy Technology Centre, Royal Observatory Edinburgh,
  Edinburgh EH9 3HJ, UK}
\affil[e]{Department of Physics, Centre for Advanced Instrumentation,
  University of Durham, Durham DH1 3LE, UK}

\authorinfo{Send correspondence to VSD -- email:
  vik.dhillon@sheffield.ac.uk, tel: +44 114 222 4528}

\begin{document} 
\maketitle

\begin{abstract}
HiPERCAM is a high-speed camera for the study of rapid variability in
the Universe. The project is funded by a \euro3.5M European Research
Council Advanced Grant. HiPERCAM builds on the success of our previous
instrument, ULTRACAM, with very significant improvements in
performance thanks to the use of the latest technologies. HiPERCAM
will use 4 dichroic beamsplitters to image simultaneously in 5 optical
channels covering the $u'g'r'i'z'$ bands. Frame rates of over 1000 per
second will be achievable using an ESO CCD controller (NGC), with
every frame GPS timestamped. The detectors are custom-made,
frame-transfer CCDs from e2v, with 4 low-noise (2.5e$^-$) outputs,
mounted in small thermoelectrically-cooled heads operated at 180 K,
resulting in virtually no dark current. The two reddest CCDs will be
deep-depletion devices with anti-etaloning, providing high quantum
efficiencies across the red part of the spectrum with no fringing. The
instrument will also incorporate scintillation noise correction via
the conjugate-plane photometry technique. The opto-mechanical chassis
will make use of additive manufacturing techniques in metal to make a
light-weight, rigid and temperature-invariant structure. First light is
expected on the 4.2\,m William Herschel Telescope on La Palma in 2017
(on which the field of view will be 10' with a 0.3"/pixel scale),
with subsequent use planned on the 10.4\,m Gran Telescopio Canarias on
La Palma (on which the field of view will be 4' with a 0.11"/pixel
scale) and the 3.5\,m New Technology Telescope in Chile.
\end{abstract}


\section{SCIENTIFIC MOTIVATION}

The next few years will be witness to a revolution in our knowledge of
the Universe with the advent of large survey facilities such as LSST,
Gaia, PLATO, SKA and Euclid. Many new variable and transient sources
will be discovered by these facilities. In the same time-frame, the
hunt will be on for electromagnetic counterparts to the gravitational
wave sources that LIGO has started to
detect\cite{2016PhRvL.116f1102A}. Time-domain astrophysics is set to
become a core activity, and detailed follow-up studies of the most
interesting objects will be essential. The world's major ground-based
telescopes provide facilities for such follow-up work, but one area is
currently poorly catered for -- high time-resolution (millisecond to
second) optical cameras.

High time-resolution probes the most extreme cosmic environments --
white dwarfs, neutron stars and black holes -- testing theories of
fundamental physics to their limits. For example, black holes and
neutron stars give us the chance to study the effects of strong-field
general relativity, and neutron stars and white dwarfs enable the
study of exotic states of matter predicted by quantum
mechanics\cite{2013Sci...340..448A}. Black holes, neutron stars and
white dwarfs are also a fossil record of stellar evolution, and the
evolution of such objects within binaries is responsible for some of
the Galaxy’s most exotic and scientifically-valuable inhabitants, such
as binary millisecond pulsars and type Ia supernovae.

One of the best ways of studying compact stellar remnants is through
their multi-colour photometric variability. The dynamical timescales
of black holes, neutron stars and white dwarfs range from milliseconds
to seconds. This means that the rotation and pulsation of these
objects, and the motion of any material in close proximity to them
(e.g. in an accretion disc), tends to occur on such short
timescales. Hence, only by observing at high speeds can the
variability of compact objects be resolved, thereby revealing a wealth
of information, such as their structure, radii, masses and emission
mechanisms\cite{2016MNRAS.459..554G}.

Observing the Universe on timescales of milliseconds to seconds is
also of benefit when studying less massive compact objects, such as
exoplanets\cite{2014MNRAS.437..475M}, brown
dwarfs\cite{2006Sci...314.1578L} and Solar System
objects\cite{2012Natur.491..566O}. Despite varying on timescales of
minutes rather than seconds, observing the eclipses and transits of
exoplanets at high time-resolution can hugely improve throughput
through the avoidance of detector readout-time, and enables the
detection of Earth-mass planets through small variations in transit
timing. Simultaneous, multi-band light curves of the transits of
exoplanets are sensitive to wavelength-dependent opacity sources in
their atmospheres, especially in low-gravity exoplanets, as will be
found by the next generation of exoplanet surveys. Within the Solar
System, high time-resolution occultation observations enable
shape/size measurements and the detection of atmospheres and ring
systems, at spatial scales (0.5 milliarcsec) only otherwise achievable
from dedicated space missions.

In this paper, we describe a new high-speed camera called HiPERCAM
that has been designed to study compact objects of all classes,
including black holes, neutron stars, white dwarfs, brown dwarfs,
exoplanets and the minor bodies of the Solar System. The resulting
data will enable us to help answer the questions: What are the
progenitors of type Ia supernovae? What are the properties of
exoplanet atmospheres? What is the equation of state of the degenerate
matter found in white dwarfs and neutron stars? What is the nature of
the flow of matter close to the event horizon of black holes? What
gravitational wave signals are likely to be detected by the next
generation of space and ground-based detectors? What are the
properties of the dwarf planets in the Kuiper belt?

\section{DESIGN}
\label{sec:design}

The design of HiPERCAM is based on our successful predecessor
instrument, ULTRACAM\cite{2007MNRAS.378..825D}, but offers a very
significant advance in performance, as shown in
Table~\ref{tab:comp} and described in the following sections.

\begin{table}[hb]
\caption{Comparison of ULTRACAM and HiPERCAM.}
\label{tab:comp}
\begin{center}
\begin{tabular}{l|l|l|l}
  & {\bf ULTRACAM} & {\bf HiPERCAM} & {\bf Notes} \\
  \hline
  Number of simultaneous colours & 3 ($u'g'+r'/i'/z'$) & 5 ($u'g'r'i'z'$) & \\
  Readout noise & 3.0\,e$^-$ at 100\,kHz & 2.5\,e$^-$ at 200\,kHz & pixel rates in kHz \\
  CCD temperature & 233\,K & 180\,K & \\
  Dark current & 360\,e$^-$/pix/hr & 6\,e$^-$/pix/hr & \\
  Longest exposure time & 30\,s & 1800\,s & \\
  Highest frame rate & 400\,Hz & 1600\,Hz & 24$\times$24 pix windows, bin 4$\times$4 \\
  Field of view on WHT & 5.1' & 10.2' & platescale 0.3''/pixel \\
  Field of view on VLT & 2.6' & -- & 0.15''/pixel (ULTRACAM) \\
  Field of view on GTC & -- & 3.9' & 0.113''/pixel (HiPERCAM) \\
  Probability of $r'=11$ comparison & 52\% & 95\% & WHT, galactic latitude = 30$^{\circ}$ \\
  Scintillation correction & No & Yes & \\
  Dummy CCD outputs & No & Yes & \\
  Deep depletion & No & Yes & \\
  QE at 700/800/900/1000\,nm & 83\%/61\%/29\%/5\% & 91\%/87\%/58\%/13\% & \\
  Fringe suppression CCDs & No & Yes & \\
  Fringe amplitude at 900\,nm & $>$10\% & $<$1\% & \\
\end{tabular}
\end{center}
\end{table}

\subsection{Optics}
\label{sec:optics}

\begin{figure}[htb]
\begin{center}
\begin{tabular}{c}
\includegraphics[width=12.5cm]{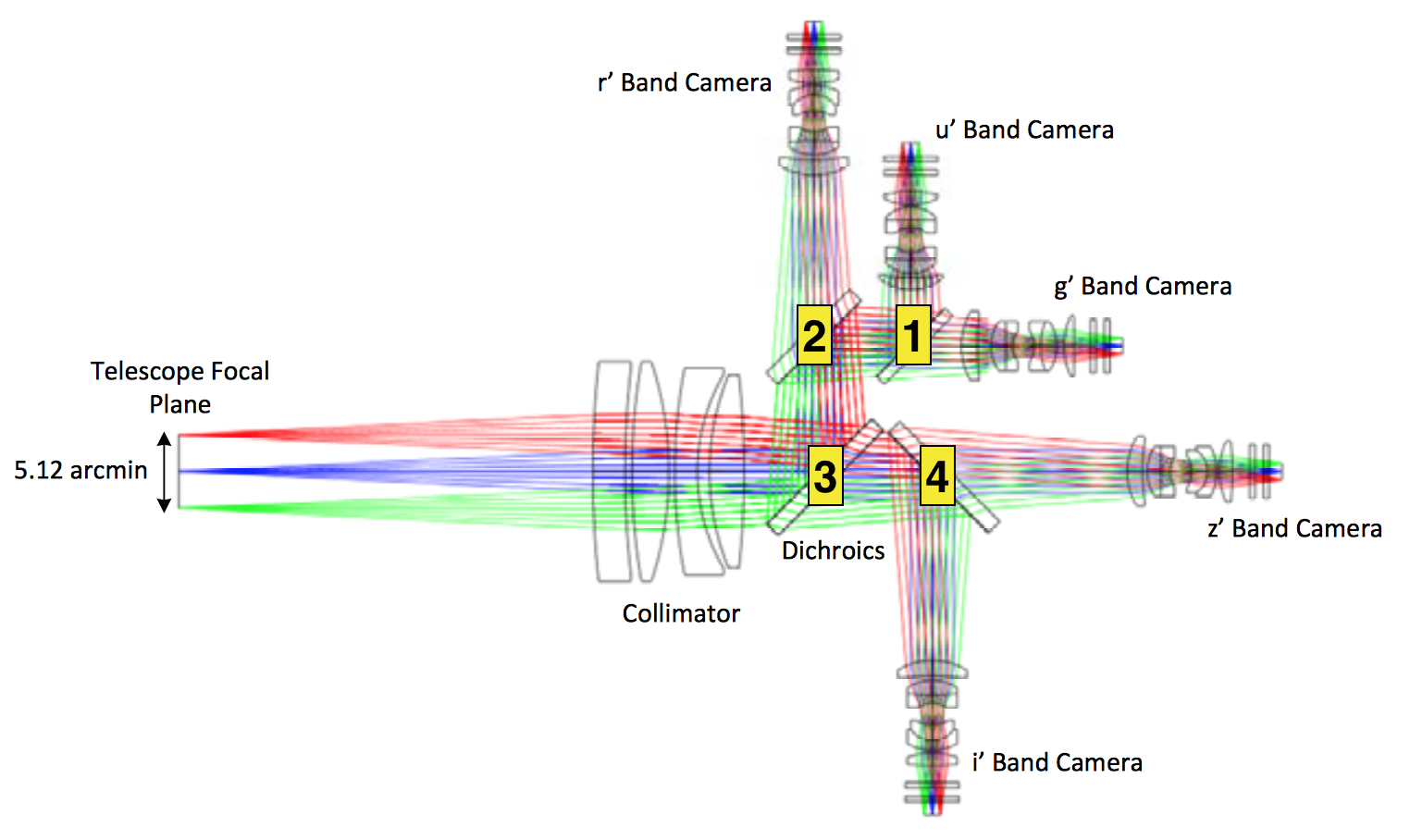}
\end{tabular}
\end{center}
\caption{Ray-trace through the HiPERCAM optics. The collimator shown
  is for the WHT/NTT. The diagram is to scale -- the largest lens in
  the collimator has a diameter of 208\,mm. The dichroics are numbered
  in ascending order of the wavelength of the cut point, as shown in
  \protect{Fig.~\ref{fig:filters}}.
\label{fig:optics}
}
\end{figure}

\begin{figure}[htb]
\begin{center}
\begin{tabular}{c}
\includegraphics[width=9.5cm]{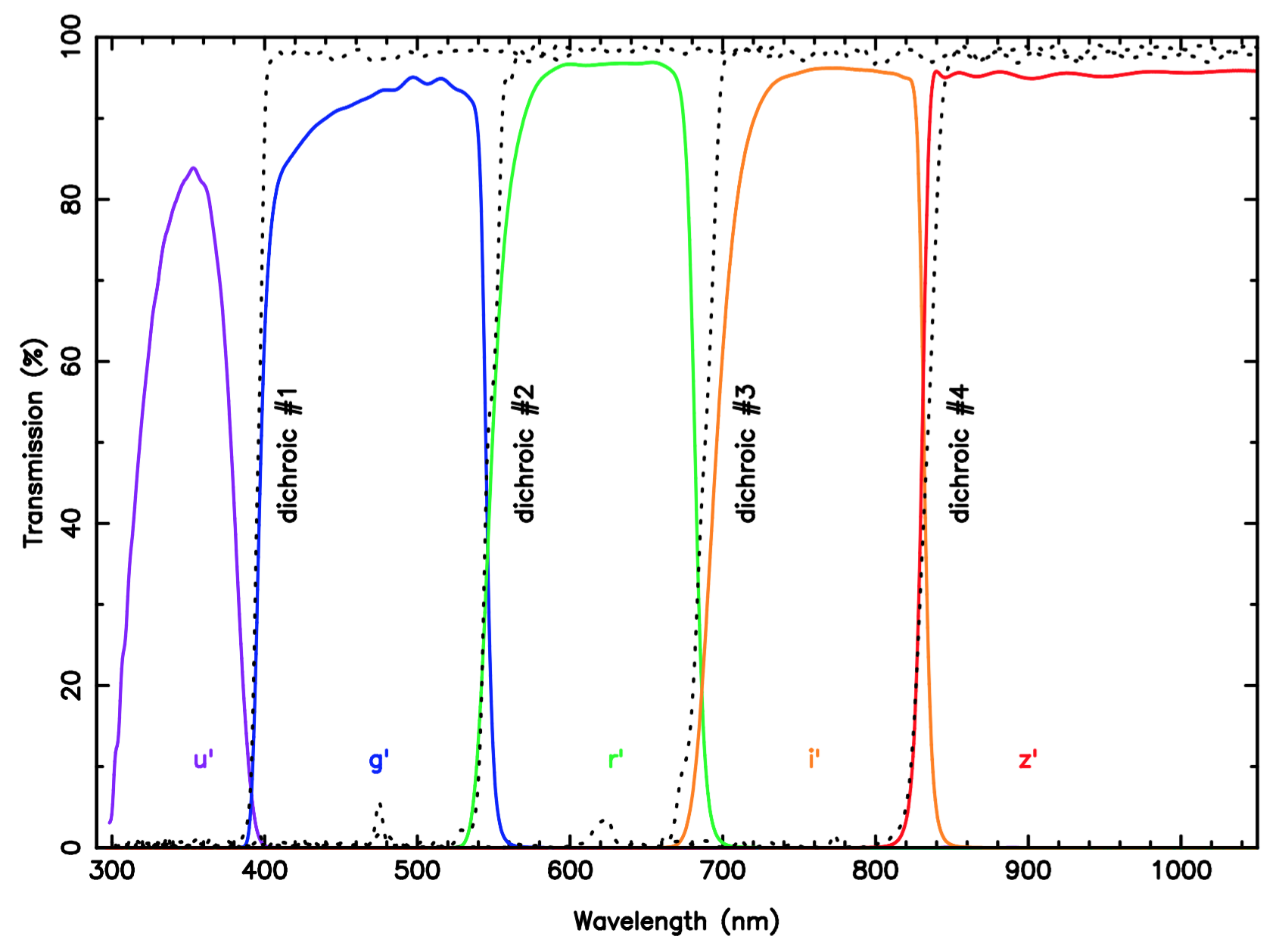}
\end{tabular}
\end{center}
\caption{Transmission profiles of the as-built HiPERCAM SDSS filters
  (solid lines) and the four dichroic beamsplitters (dotted lines).
\label{fig:filters}
}
\end{figure}

The HiPERCAM optical design is shown in Fig.~\ref{fig:optics}. Light
from the telescope is first collimated and then split into five arms
using four dichroic beamsplitters. The beam in each arm then passes
through a re-imaging camera, which focuses the light through a
bandpass filter and cryostat window onto a CCD. The bandpasses of the
five arms are defined by SDSS $u'g'r'i'z'$ filters
(Fig.~\ref{fig:filters}), although narrow-band filters that lie within
these five bandpasses can also be used. The collimator shown in
Fig.~\ref{fig:optics} has been designed for use on both the 4.2\,m
William Herschel Telescope (WHT) on La Palma and the 3.5\,m New
Technology Telescope (NTT) on La Silla, giving a platescale of
0.3''/pixel and 0.35''/pixel, respectively, and a field of view of
10.2' and 12.0', respectively. On the WHT, HiPERCAM will have a 95\%
chance of observing a comparison star of magnitude $r'=11$ in the
field of view (at a galactic latitude of 30$^{\circ}$) for
differential photometry.  A second collimator for use on the 10.4\,m
Gran Telescopio de Canarias (GTC) has also been designed, giving a
platescale of 0.113''/pixel and a field of view of 3.9'.  When
switching between telescopes, only the collimator needs to be changed
-- no other optics need to be added, removed or re-aligned.

\begin{figure}[htb]
\begin{center}
\begin{tabular}{c}
\includegraphics[width=12.0cm]{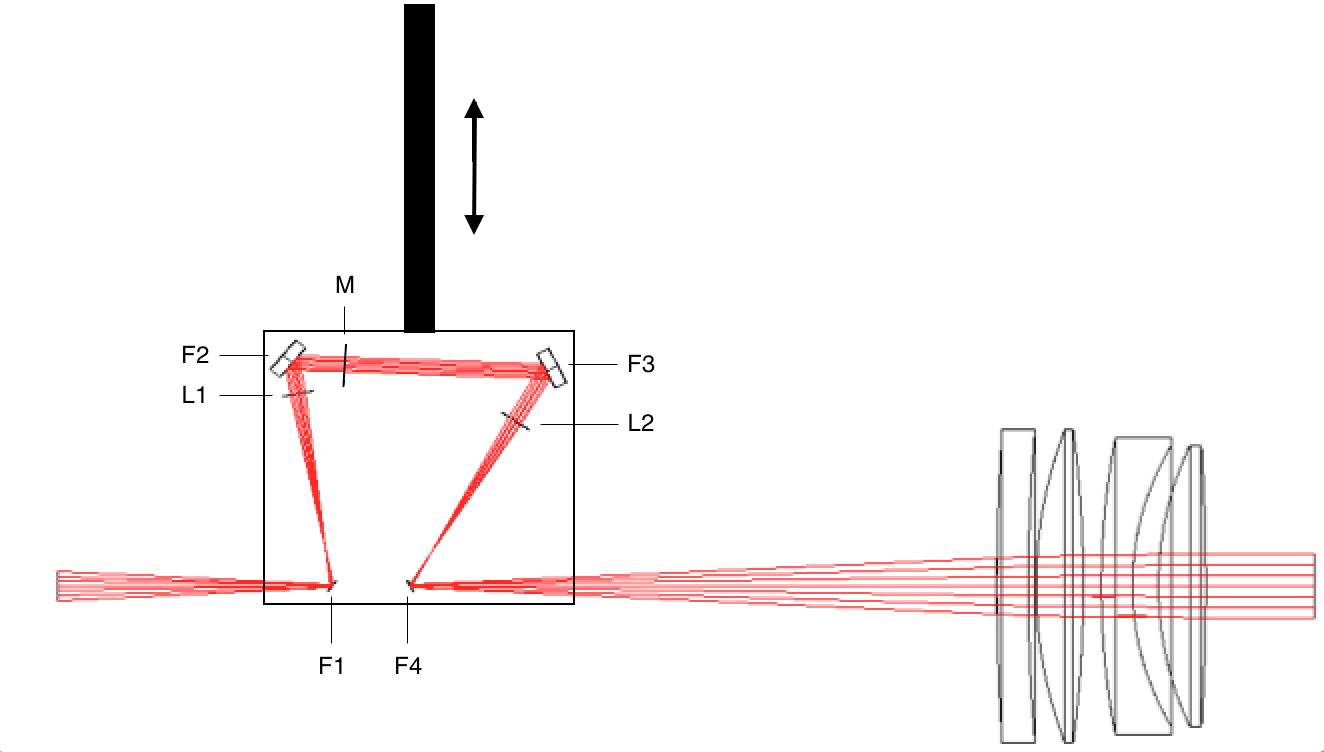}
\end{tabular}
\end{center}
\caption{Schematic showing one of the two scintillation corrector
  pick-offs, which will be mounted in the interface collar between
  HiPERCAM and the telescope. All of the optical components in each
  pick-off will be mounted on a single base that slides in and out of
  the beam on a linear translation stage. Converging light from the
  telescope is picked off by a flat (F1), collimated (L1), folded by
  another flat (F2), masked (M), folded by another flat (F3),
  re-imaged (L2) and then re-introduced by a final flat (F4) into
  HiPERCAM with an identical focal ratio and at an identical focal
  point as an unperturbed beam.
  \label{fig:cpp}
}
\end{figure}

HiPERCAM also incorporates a scintillation corrector, which
exploits the Conjugate-Plane Photometry technique
\cite{2011MNRAS.411.1223O,2015MNRAS.452.1707O}. The optical
layout of the scintillation corrector is depicted in
Fig.~\ref{fig:cpp}. Light from the target star is picked off by a flat
lying just above the focal plane of the telescope, collimated and then
masked at the conjugate altitude of the turbulent layer. The beam is
then re-imaged in such a way that a second flat lying just below the
focal plane of the telescope returns the beam into HiPERCAM with an
identical focal ratio and focal point as an unperturbed beam. Hence,
as far as HiPERCAM is concerned, the pick-off system is “transparent”.
Two such pick-offs will be required, one for the target star and
another for the comparison star.

\subsection{Detectors}
\label{sec:detectors}

\begin{figure}[htb]
\begin{center}
\begin{tabular}{c}
\includegraphics[width=7.5cm]{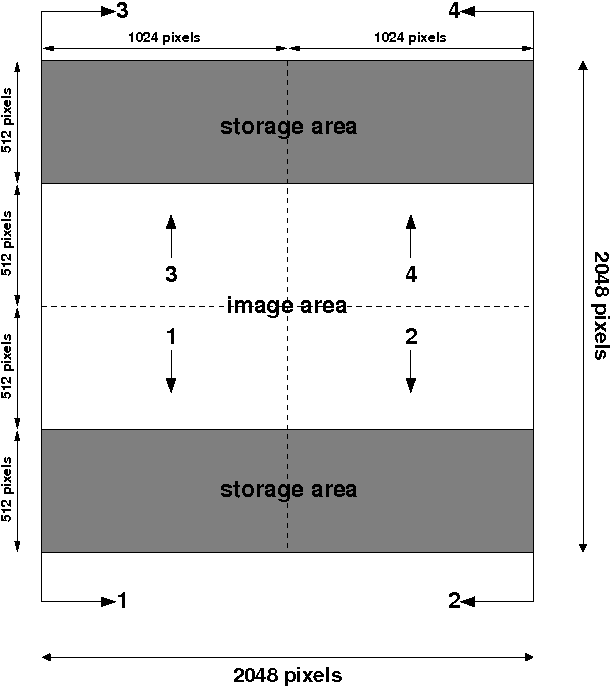}
\end{tabular}
\end{center}
\caption{Schematic of one of the HiPERCAM e2v CCD231-42 detectors.
\label{fig:ccds}
}
\end{figure} 

HiPERCAM will employ 5 custom-designed CCD231-42 CCDs from e2v. Each
CCD is a frame-transfer device with 15\,$\mu$m pixels and 4 low-noise
outputs, with one output located on each corner. The CCDs have a
format of 2048$\times$2048 pixels, where the upper 2048$\times$512 and
lower 2048$\times$512 pixels are masked off as frame-transfer storage
regions, providing an image area of 2048$\times$1024 pixels. Hence
each CCD output processes a quadrant of format 1024$\times$512 pixels,
as shown in Fig.~\ref{fig:ccds}.

The HiPERCAM CCD231-42 CCDs are back-illuminated and thinned for high
quantum efficiency (QE). The $u'$ CCD has an Astro Broadband
anti-reflection (AR) coating, and the $g', r', i'$ and $z'$ CCDs have
the Astro Multi-2 AR coating. The $u', g'$ and $r'$ CCDs are
manufactured from standard silicon, whereas the $i'$ and $z'$ CCDs
are manufactured from deep depletion silicon to maximise red QE. The
$i'$ and $z'$ CCDs also incorporate e2v's fringe suppression
process. This breaks the interference condition by introducing
irregularities in the back surface of the device, reducing fringing to
approximately the same level as the flat-field noise
\cite{tulloch2005}.

To maximise readout speed and minimise noise, the HiPERCAM CCDs are
equipped with: low noise outputs of 2.5\,e$^-$ at 200 kHz; dummy
outputs to eliminate pickup noise; fast row (vertical) clocking of
8\,$\mu$s/row; fast serial (horizontal) clocking of 0.1\,$\mu$s/pixel;
the ability to clock the serial register of each quadrant
independently in order to enable efficient windowing
modes\cite{2007MNRAS.378..825D}; two-phase image and storage clocks to
minimise the frame-transfer~time.

We have decided to use non inverted-mode (NIMO) rather than
inverted-mode (AIMO) devices in HiPERCAM, for three reasons. First,
NIMO devices can be clocked more quickly. Second, although both AIMO
and NIMO devices can have similar mean dark current specifications at
their optimum operating temperatures, our experience with ULTRASPEC
(NIMO\cite{2014MNRAS.444.4009D}) and ULTRACAM
(AIMO\cite{2007MNRAS.378..825D}) is that the dark current in NIMO
devices is much more uniformly distributed than in AIMO CCDs; the dark
current in the latter is predominantly in the form of numerous hot
pixels which do not subtract well using dark frames, making exposures
longer than $\sim30$\,s undesirable. Third, and most importantly, we
chose to use NIMO devices in HiPERCAM because we wish to use
deep-depletion silicon in the red CCDs to maximise QE and this in
incompatible with inverted-mode operation.

\begin{figure}[htb]
\begin{center}
\begin{tabular}{c}
\includegraphics[width=6.9cm]{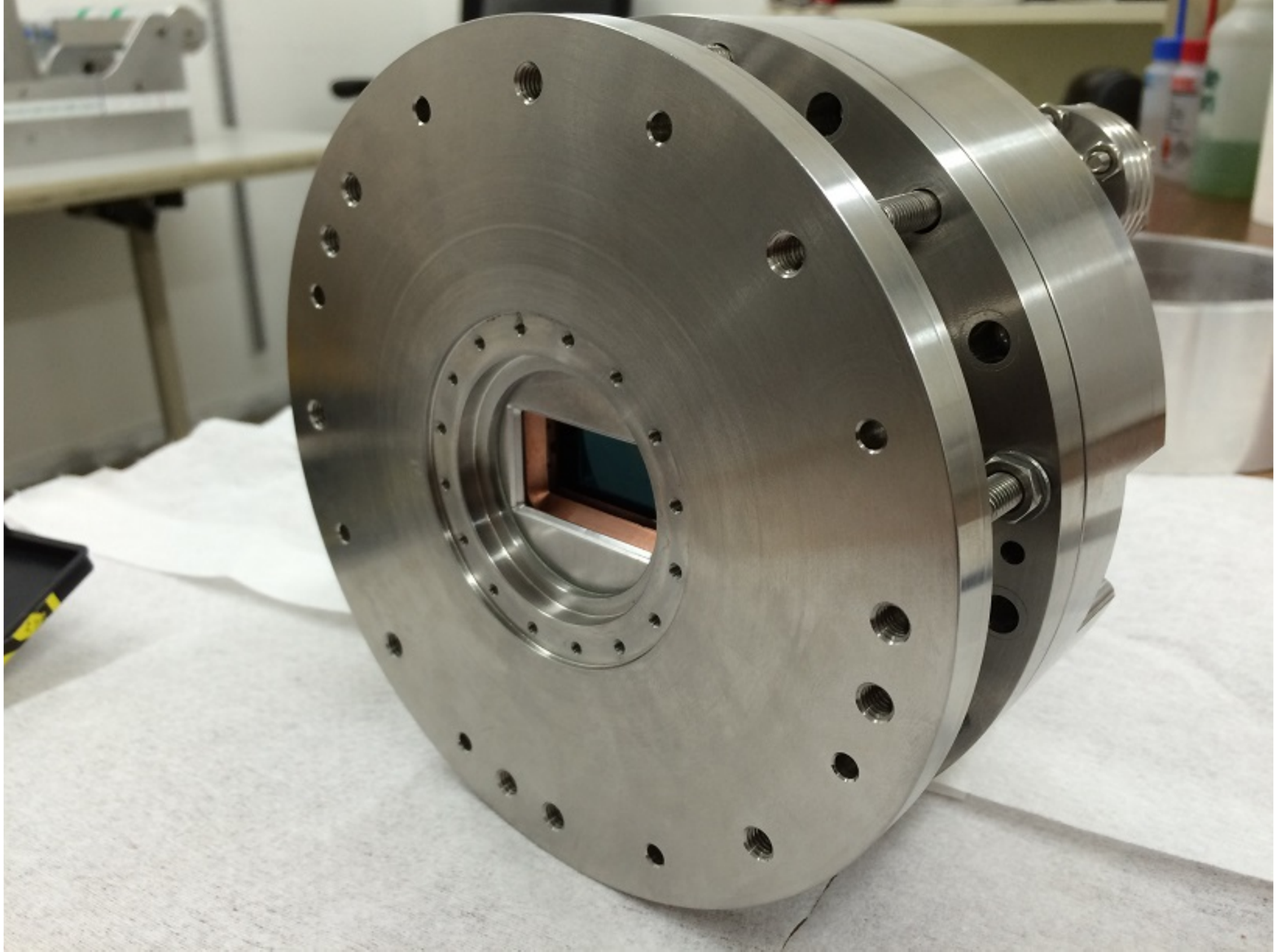}
\end{tabular}
\end{center}
\caption{Photograph of one of the TEC-cooled HiPERCAM CCD heads. The
  diameter of the head is 160\,mm and the weight is approximately
  5\,kg.
\label{fig:head}
}
\end{figure}

The consequence of selecting NIMO devices for HiPERCAM is that the
detectors need to be cooled to 180\,K or below to ensure that the dark
current is less than 360\,e$^-$/pixel/hr, which corresponds to
10\%\ of the faintest sky level we will record with HiPERCAM (set by
observations in the $u'$-band in dark time on the GTC). Cooling to
180\,K will therefore ensure dark current is always a negligible noise
source in HiPERCAM. To achieve 180\,K, we looked at a number of
cooling options. We eliminated liquid nitrogen as impractical:
incorporating five liquid-nitrogen cryostats would make HiPERCAM
large, heavy, and very time-consuming to fill each night (automatic
filling or continuous flow systems are not an option given that
HiPERCAM will be a visitor instrument).  We also eliminated
closed-cycle Joule-Thomson coolers, such as the CryoTiger, due to the
impracticality of housing up to 5 compressors at the telescope and
passing up to 10 stainless-steel braided gas lines through the cable
wrap. We seriously considered Stirling coolers, but were concerned
about the vibrations that they induce in the telescope. These
vibrations can be reduced through careful choice of cooler and the use
of complex, bulky anti-vibration mounts \cite{2013arXiv1311.0685R},
but even with these precautions in place, it would have been difficult
to persuade the potential host telescopes to accept HiPERCAM with
Stirling coolers due to the residual vibrations. Eventually, following
extensive prototyping and testing, we decided to use thermo-electric
(peltier) coolers (TECs), which are the simplest, cheapest, lightest
and most compact of all of the cooling options. By using two
five-stage TECs in conjunction with careful control of the parasitic
heat load on the detector, we have been able to build a CCD head that
can achieve the target detector temperature of 180\,K. The heat
generated by the TECs is extracted using a 278\,K water-glycol cooling
circuit. The interior of the head has also been carefully designed to
maximise the lifetime of the vacuum -- we anticipate a hold time of
years. A photograph of one of the heads~is~shown~in~Fig.~\ref{fig:head}.

\subsection{Mechanics}
\label{sec:mechanics}

\begin{figure}[b]
\begin{center}
\begin{tabular}{c}
\includegraphics[width=8.9cm]{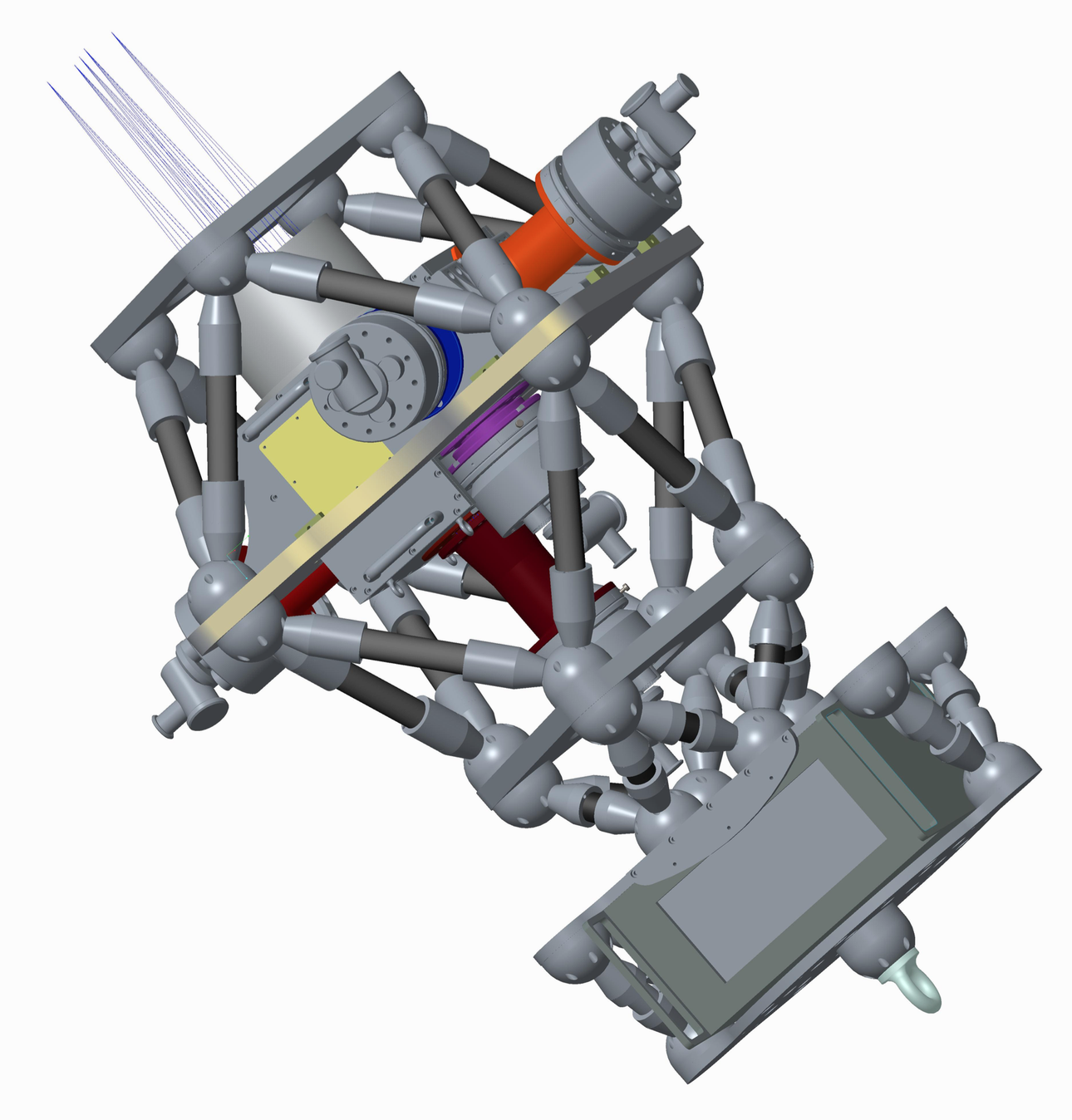}
\end{tabular}
\end{center}
\caption{CAD image of the HiPERCAM opto-mechanical assembly. The length and
weight of the instrument are approximately 1.25\,m and 200\,kg, respectively.
\label{fig:mech}
}
\end{figure}

The HiPERCAM opto-mechanical chassis (Fig.~\ref{fig:mech}) is a triple
octopod composed primarily of 3 large aluminium plates connected by
carbon fibre struts, and incorporating some components made using
additive manufacturing techniques in metal (titanium and invar). The
design results in a stiff, compact (1.25\,m long), light-weight
(200\,kg) and open structure, which is relatively insensitive to
temperature variations. These characteristics are essential for
visitor instruments, as they make HiPERCAM easy to maintain, transport
and mount/dismount at the telescope.

The dichroics, lens barrels, filters and CCDs are housed in/on an
aluminium hull, which forms a sealed system to light and dust. The
hull is affixed to the large central aluminium plate shown in
Fig.~\ref{fig:mech}. The top plate connects the instrument to the
telescope, and the CCD controller is mounted on the bottom plate.

\subsection{Data acquisition system}
\label{sec:das}

The HiPERCAM data acquisition system is designed to be detector
limited, i.e. the throughput of data from the output of the CCDs to
the hard disk on which it is eventually archived is always greater
than the rate at which the data comes off the CCDs. This means that
the instrument is capable of running continuously all night at its
maximum data rate without ever having to pause for archiving of data.

\begin{figure}[htb]
\begin{center}
\begin{tabular}{c}
\includegraphics[width=11.0cm]{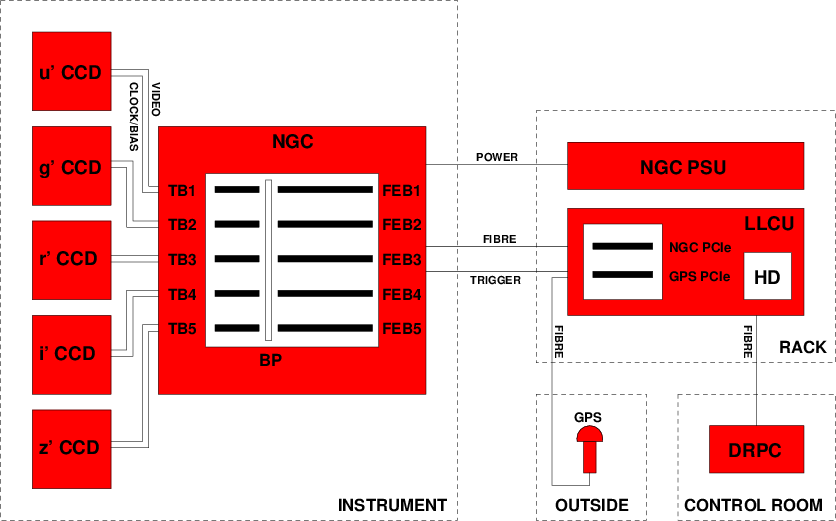}
\end{tabular}
\end{center}
\caption{Schematic of the HiPERCAM data acquisition system -- see text
  for details.  
\label{fig:das}
}
\end{figure}

Fig.~\ref{fig:das} shows a schematic of the HiPERCAM data acquisition
system. The architecture is similar to that developed for
ULTRACAM\cite{2007MNRAS.378..825D}, but uses much more up-to-date,
faster hardware. The heart of the system is a European Southern
Observatory (ESO) New General detector Controller
(NGC)\cite{2009Msngr.136...20B}. The HiPERCAM NGC is composed of a
six-slot housing containing 5 Front-End Basic (FEB) Boards and 5
Transition Boards (TB). Each FEB is connected to a back plane (BP) and
is accompanied by a TB, which handles all external connections to the
FEB. Each TB is connected to a CCD head via a single cable, carrying
both the clocks/biases from the clock/bias-driver on the associated
FEB and the CCD video signal to the four ADC channels on the FEB. The
NGC is located on the instrument itself (see Fig.~\ref{fig:mech}) in
order to minimise the length of the cables running to each CCD head
for noise reduction.

The NGC is powered by a separate Power Supply Unit (PSU), which is
located in an electronics rack, approximately 2--5\,m away from the
instrument. The rack also contains the Linux Local Control Unit (LLCU)
for the NGC, which is a linux PC provided by ESO containing the NGC
Peripheral Computer Interconnect Express (PCIe) card. The NGC PCIe
card is connected to the NGC via fibre, through which it is possible
to control the NGC and receive data. In addition to the system disk,
the LLCU contains a large-capacity hard disk (HD) on which the raw CCD
data are written.

The LLCU also contains a GPS PCIe card, which accepts two inputs. The
first input is a trigger that is generated by the NGC whenever an
exposure finishes. This trigger will cause the GPS card to write a
timestamp to its FIFO (First In, First Out buffer), which is then
written to the header of the corresponding CCD frame. The second input
is a GPS signal from an external antenna. The external antenna will be
connected to the GPS card in the LLCU via a fibre-converter that
enables long cable runs to outside the dome and isolates the telescope
from lightning strikes.

A second linux PC, referred to as the Data Reduction PC (DRPC) in
Fig.~\ref{fig:das}, will be located in the telescope control room.
This PC will run, amongst other things, the data reduction pipeline,
the instrument-control GUI, the data logger and the target acquisition
tool. It will be connected to the LLCU via fibre ethernet.

\subsection{Performance}
\label{sec:performance}

\begin{figure}[htb]
\begin{center}
\begin{tabular}{c}
\includegraphics[width=9.0cm]{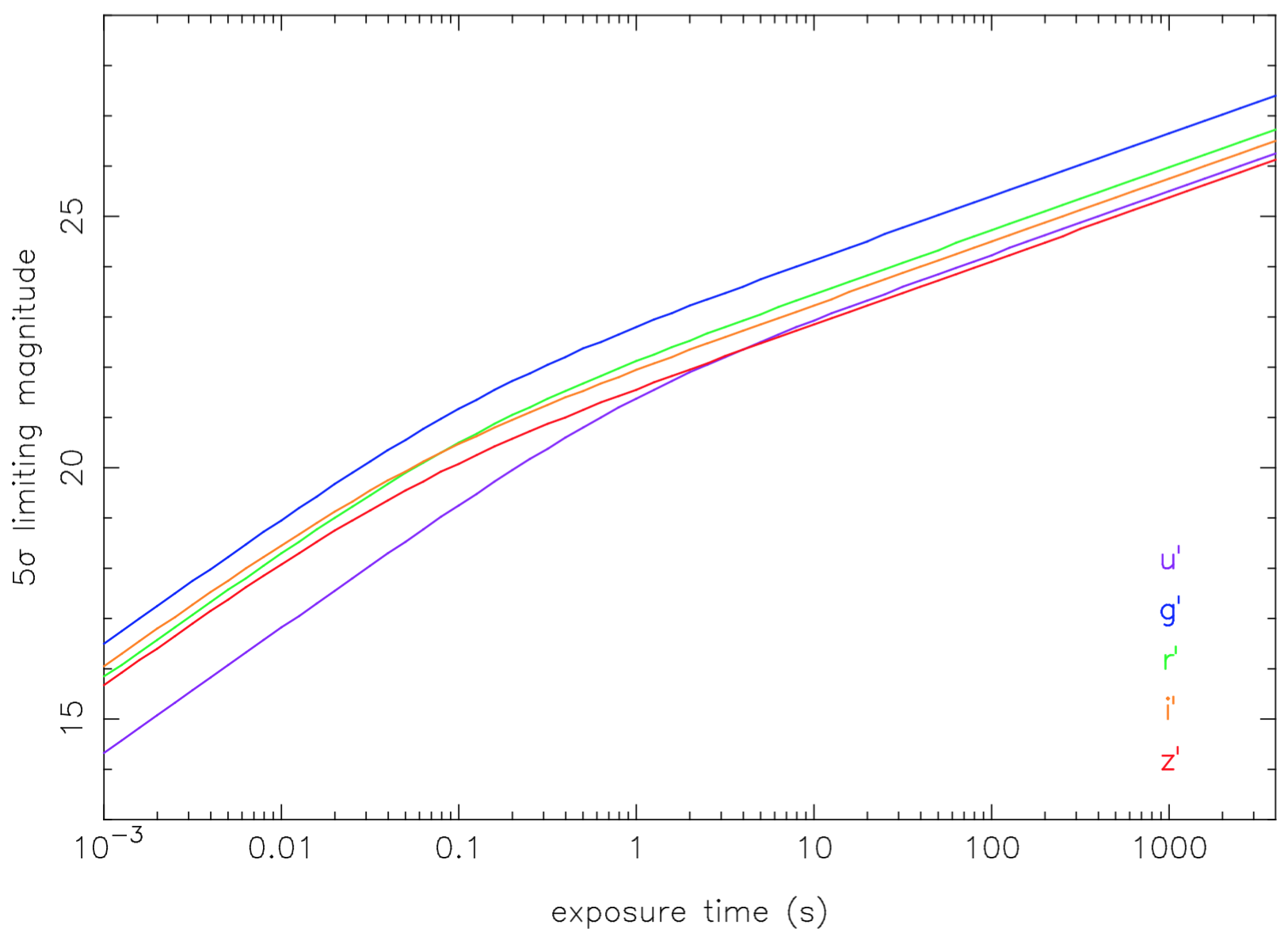}
\end{tabular}
\end{center}
\caption{Limiting magnitudes (5$\sigma$) of HiPERCAM on the GTC as a
  function of exposure time. The purple, blue, green, orange and red
  curves show the results for the $u'$, $g'$, $r'$, $i'$ and $z'$
  filters, respectively.
\label{fig:snr}
}
\end{figure}

Fig.~\ref{fig:snr} shows the predicted 5σ limiting magnitudes
achievable with HiPERCAM on the GTC as a function of exposure
time. The calculations are based on the measured zeropoints for
ULTRACAM on the WHT\cite{2007MNRAS.378..825D}, but scaled for: the
increased telescope aperture of the GTC; the extra reflection off the
Nasmyth flat at the GTC; the improved dichroics and AR coatings used
in HiPERCAM; the superior red QEs of the CCDs in HiPERCAM. The
calculations also assume dark moon, observing at the zenith and seeing
of 0.8''. It can be seen that it should be possible to achieve a
limiting magnitude of $g'\sim 16.5$ in a single 0.001\,s exposure,
$g'\sim 23$ in 1\,s, and $g'\sim 27$ in 1800\,s.

Due to the fast vertical clocking time and split-frame transfer
architecture of the CCDs in HiPERCAM, the dead time between exposures
in standard readout mode will be only 0.004\,s, a factor of 6 times
less than ULTRACAM. However, substantially lower dead times will be
available in drift mode\cite{2007MNRAS.378..825D}, where the windows
are positioned on the border between the image and storage areas and,
instead of vertically clocking the entire image area into the storage
area, only the window is clocked into the (top of) the storage area.
In drift mode, we predict frame rates of 1600\,Hz using four
$24\times24$ pixel windows, binned $4\times4$, for example. This is a
factor of 4 times quicker than ULTRACAM, and is primarily due to the
much faster horizontal clocking provided by the NGC and the CCDs used
in HiPERCAM.

\section{CONCLUSIONS}

We have described the scientific motivation, design and predicted
performance of HiPERCAM. The instrument offers a significant advance
on what has come before in the field of high time-resolution optical
instrumentation. First light is planned for 2017 on the 4.2\,m WHT,
with subsequent use planned on the 10.4\,m GTC later that year.

\acknowledgments
 
HiPERCAM is funded by the European Research Council under the European
Union's Seventh Framework Programme (FP/2007-2013) under ERC-2013-ADG
Grant Agreement no. 340040 (HiPERCAM). We would like to thank our
major suppliers for their assistance with this project: Amptown, Asahi
Spectra, e2v, ESO, Fluidic, Huber, Labtex, Meerstetter, MKS
Instruments, Newport, Rocky Mountain Instruments, Staubli, Steatite,
Titan Enterprises, Universal Cryogenics.

\bibliography{hipercam_spie} 
\bibliographystyle{spiebib} 

\end{document}